\documentclass[12pt]{article}

\usepackage{natbib, array}

\usepackage{amsmath, amsbsy, epsfig, epsf, psfrag, booktabs, graphicx, amssymb, enumerate, amsthm}
\usepackage{hypernat}
\usepackage{multicol}

\usepackage[margin=1in]{geometry}

\usepackage[pagewise]{lineno}
\linenumbers
    \newcommand*\patchAmsMathEnvironmentForLineno[1]{%
      \expandafter\let\csname old#1\expandafter\endcsname\csname #1\endcsname
      \expandafter\let\csname oldend#1\expandafter\endcsname\csname end#1\endcsname
      \renewenvironment{#1}%
         {\linenomath\csname old#1\endcsname}%
         {\csname oldend#1\endcsname\endlinenomath}}%
    \newcommand*\patchBothAmsMathEnvironmentsForLineno[1]{%
      \patchAmsMathEnvironmentForLineno{#1}%
      \patchAmsMathEnvironmentForLineno{#1*}}%
    \AtBeginDocument{%
    \patchBothAmsMathEnvironmentsForLineno{equation}%
    \patchBothAmsMathEnvironmentsForLineno{align}%
    \patchBothAmsMathEnvironmentsForLineno{flalign}%
    \patchBothAmsMathEnvironmentsForLineno{alignat}%
    \patchBothAmsMathEnvironmentsForLineno{gather}%
    \patchBothAmsMathEnvironmentsForLineno{multline}%
    }

\newcommand{\dif}{\mathrm{d}}%
\newcommand{\Var}{\mathrm{Var}}%
\newcommand{\Cov}{\mathrm{Cov}}%

\newtheorem{thm}{Theorem}

\graphicspath{{./figs/}}

\begin{document}

\markright{
}
\markboth{\hfill{\footnotesize\rm S. Chiou, J. Kim AND J. Yan} \hfill}
{\hfill {\footnotesize\rm AFT Models with GEE} \hfill}

\renewcommand{\thefootnote}{}
$\ $\par


\fontsize{10.95}{14pt plus.8pt minus .6pt}\selectfont
\vspace{0.8pc}
\centerline{\large\bf Semiparametric Multivariate Accelerated Failure Time Model }
\vspace{2pt}
\centerline{\large\bf with Generalized Estimating Equations}
\vspace{.4cm}
\centerline{Sy Han Chiou, Junghi Kim, and Jun Yan}
\vspace{.4cm}
\centerline{\it University of Conecticut, University of Minnesota }
\centerline{\it and University of Connecticut Health Center}
\vspace{.55cm}
\fontsize{9}{11.5pt plus.8pt minus .6pt}\selectfont



\begin{quotation}
\noindent {\it Abstract:}
The semiparametric accelerated failure time model is not as widely used
as the Cox relative risk model mainly due to computational difficulties.
Recent developments in least squares estimation and induced
smoothing estimating equations provide promising tools to make
the accelerate failure time models more attractive in practice.
For semiparametric multivariate accelerated failure time models, 
we propose a generalized estimating equation approach to account for 
the multivariate dependence through working correlation structures.
The marginal error distributions can be either identical as in
sequential event settings or different as in parallel event settings.
Some regression coefficients can be shared across margins as needed.
The initial estimator is a rank-based estimator with Gehan's weight,
but obtained from an induced smoothing approach with computation ease.
The resulting estimator is consistent and asymptotically normal,
with a variance estimated through a multiplier resampling method.
In a simulation study, our estimator was up to three
times as efficient as the initial estimator, especially with
stronger multivariate dependence and heavier censoring percentage.
Two real examples demonstrate the utility of the proposed method.

\vspace{9pt}
\end{quotation}\par

\fontsize{10.95}{14pt plus.8pt minus .6pt}\selectfont

\section{Introduction}
\label{sect:intr}

Multivariate failure times are frequently encountered in
biomedical research where failure times are clustered.
For example, a diabetic retinopathy study assessed the 
efficacy of a laser treatment on decelerating vision loss, 
measured by time to blindness in the left eye and in the 
right eye from the same patient with diabetes \citep{DRS:1976}; 
a colon cancer study evaluated the treatment effects on prolonging
the time to tumor recurrence and time to death \citep{Lin:cox:1994}.
The failure times within the same cluster are associated.
Even though the primary interest most often lies in the
marginal effects of covariates on the failure times,
accounting for the within-cluster dependence may lead to
more efficient regression coefficient estimators.
For non-censored multivariate data, the generalized estimating
equations (GEE) approach \citep{Lian:Zege:long:1986} has become an
important piece in statisticians' toolbox for marginal regression.
For censored multivariate failure times, the marginal accelerated
failure time (AFT) model is a counterpart of the marginal model.
This paper aims to develop a GEE approach to make inferences
for multivariate AFT models, taking advantage of recent
developments on AFT models with least squares and induced smoothing.

A semiparametric AFT model is a linear model for the logarithm 
of the failure times with error distribution unspecified.
A nice interpretation is that the effect of a covariate
is to multiply the predicted failure time by some constant.
It provides an attractive alternative to the
popular relative risk model \citep{Cox:regr:1972}.
Three main classes of estimator exist for univariate AFT models.
The Buckley--James (BJ) estimator extends the least squares principle 
to accommodate censoring through an expectation--maximization (EM) 
algorithm which iterates between imputing the censored failure times 
and least squares estimation \citep{Buck:Jame:line:1979}.
Despite the nice asymptotic properties 
\citep{Rito:esti:1990, Lai:Ying:larg:1991}, the BJ estimator may 
be hard to get as the EM algorithm may not converge.
Further, the limiting covariance matrix is difficult to estimate
because it involves the unknown hazard function of the error term.
The second class is the rank-based estimator motivated by 
inverting the weighted log-rank test \citep{Pren:line:1978}.
Its asymptotic properties has been rigorously studied by 
\citet{Tsia:esti:1990} and \citet{Ying:larg:1993}.
Due to lack of efficient and reliable computing algorithm, the 
rank-based estimator has not been widely used in practice until 
recently, with numerical strategies for drawing inference developed
by \citet{Huan:cali:2002} and \citet{Stra:acce:2005}.
The third class is obtained by minimizing an inverse probability
of censoring weighed (IPCW) loss function \citep{Robi:Rotn:reco:1992}.
The IPCW estimator is easy to compute, consistent and asymptotically normal
\citep{Zhou:$m$-:1992, Stut:cons:1993, Stut:dist:1996}, but it requires
correct specification of the conditional censoring distribution and
overlapping of the supports of the censoring time and the failure time.

More recent works have led to a promising perspective on
bringing AFT models into routine data analysis practice.
For rank-based inference, \citet{Jin:Lin:Wei:Ying:rank:2003}
proposed a linear programming approach, exploiting the fact that the
weighted rank estimating equation is the gradient of an objective
function which can be readily solved by linear programming.
Variances of the estimators are obtained from a resampling method.
A computationally more efficient approach for rank-based
inference with Gehan's weight \citep{Geha:gene:1965} is the
induced smoothing procedure of \citet{Brow:Wang:indu:2007}.
This approach is an application of the general induced smoothing
method of \citet{Brow:Wang:stan:2005}, where the discontinuous
estimating equations are replaced with a smoothed version, whose
solutions are asymptotically equivalent to those of the former.
The smoothed estimating equations are differentiable, which
facilitates rapid numerical solution and sandwich variance estimator.
\citet{Jin:Lin:Ying:on:2006} suggested an iterative least-squared
procedure that starts from a consistent and asymptotically
normal initial estimator such as the one obtained from the 
rank-based method of \citet{Jin:Lin:Wei:Ying:rank:2003}.
The resulting estimator is consistent and asymptotically normal, with
variance estimated from a multiplier resampling approach.

For multivariate AFT models, \citet{Jin:Lin:Ying:rank:2006}
developed rank-based estimating equations that are solved
via linear programming for marginal regression parameters.
\citet{John:Stra:indu:2009} extended the induced smoothing
approach for a rank-based estimator with Gehan's weight
to the case of clustered failure times and showed that the
smoothed estimates perform as well as those from the best
competing methods at a fraction of the computational cost.
\citet{Jin:Lin:Ying:on:2006} considered their least squares
method with marginal models for multivariate failure times.
All these approaches used independent working model and
left the within-cluster dependence structure unspecified.
\citet{Li:Yin:gene:2009} developed a generalized method of 
moments approach for rank-based estimator using the quadratic
inference function approach \citep{Qu:Lind:Li:impr:2000}
to incorporate within-cluster dependence.
\citet{Wang:Fu:rank:2011} incorporated within-cluster ranks
for the Gehan type estimator with the aid of induced smoothing.
To the best of our knowledge, little work has been done to
extend the GEE approach to the setting of multivariate AFT models
except a technical report \citep{Horn:Hame:comb:1996},
where the BJ estimator was combined with GEE.
Nevertheless, having no access to recent advances on AFT models,
they did not solve the convergence problems, and their asymptotic
variance estimator formula could not be easily computed because it
depends on the derivatives of imputed failure times with
respect to regression parameters, which might explain 
their overestimation of the variance.

We propose an iterative GEE procedure to account for multivariate
dependence through a working covariance or weight matrix.
This method has the same spirit as GEE in that misspecification
of the working covariance matrix does not affect the consistency
of the parameter estimator in the marginal AFT models; when
the working covariance is close to the unknown truth, the
estimator has higher efficiency than that from working
independence as used in \citet{Jin:Lin:Ying:on:2006}.
Our initial estimator is the computationally efficient, rank-based
estimator from \citet{John:Stra:indu:2009}, whose consistency and
asymptotic normality is inherited by the resulting GEE estimator.
We develop methods for cases where all marginal distributions are
identical and for cases where at least two margins are different.
Regression coefficients can be the same or partially the same
across margins as needed.

The rest of the article is organized as follows.
The semiparametric multivariate accelerated failure time model 
and the notation are introduced in Section~\ref{sect:maft}.
In Section~\ref{sect:gee}, we propose an iterative GEE procedure
to update a consistent and asymptotically normal initial estimator 
and present asymptotic properties of our estimator. 
A large scale simulation study is reported in Section~\ref{sect:simu}
to assess the properties of the proposed estimator.
The proposed methods are illustrated with the two 
aforementioned real applications in Section~\ref{sect:appl}.
In particular, some new findings are 
reported in analyzing the diabetic retinopathy study.
A discussion concludes in Section~\ref{sect:disc}.
The sketch of proofs are relegated to the appendix.

\section{Multivariate Accelerated Failure Time Model}
\label{sect:maft}

There are two types of multivariate failure times depending
on whether the multiple events are parallel or sequential.
The difference between the two types is that the dimension is
fixed for parallel data while random for sequential data.
In a regression model, we generally have different covariates
and different coefficients at each margin for parallel data.
For sequential data, however, some or all covariates and
covariate coefficients may be the same across margins.
In general, it is desirable to allow some of the regression 
coefficients to be shared across margins as needed.
We develop the methodology for parallel data for notational simplicity
but comment when appropriate on how to adapt to sequential data.

Consider a random sample formed by $n$ clusters.
For parallel data, all clusters are of size $K$
while for sequential data, cluster $i$ may have size $K_i$.
For ease of notation, assume at the moment that
the cluster sizes are all equal to $K$.
For $i = 1, \cdots,n$ and $k = 1, \cdots, K$, let
$T_{ik}$ and $C_{ik}$ be, respectively, the log-transformed
failure time and censoring time for margin $k$ in cluster $i$.
Let $Y_{ik} = \min(T_{ik}, C_{ik})$ and $\Delta_{ik} = I(T_{ik} < C_{ik})$.
We stack $Y_{ik}$, $T_{ik}$,  $C_{ik}$, and $\Delta_{ik}$, 
$k = 1, \ldots, K$, to form $K\times 1$ vector $Y_{i}$, $T_{i}$, 
$C_{i}$, and $\Delta_i$, respectively.
Let $X_{i} = (X_{i1}, \ldots, X_{iK})^{\top}$ be a $K \times p$ 
covariate matrix, with the $k$th row denoted by $X_{ik}$.
The observed data are independent and identically
distributed copies of $\{Y, \Delta, X\}$:
$\{(Y_{i}, \Delta_{i}, X_{i}): i = 1, \ldots, n\}$.
We assume that $T_{i}$ and $C_{i}$ are conditionally independent
given $X_i$.

Our multivariate accelerated failure time model is
\begin{equation}
\label{equ:maft}
T_{i} = X_{i} \beta + \epsilon_{i},
\end{equation}
where $\beta$ is a $p \times 1$ vector of regression coefficients,
and $\epsilon_{i} = (\epsilon_{i1}, \ldots, \epsilon_{iK})^{\top}$ is a
random error vector with an unspecified multivariate distribution.
This formulation accommodates margin-specific regression
coefficients, in which case,  $\beta$ is a stack of all
marginal coefficients, and $X_{i}$ is a block diagonal matrix.
The error vectors $\epsilon_{i}$'s, $i = 1, \ldots, n$,
are independent and identically distributed.
For parallel data, the $K$ marginal distributions can be all different,
while for sequential data, the number of unique marginal distributions
may be smaller or even one as in a recurrent event setting.

With right censoring, \citet{Buck:Jame:line:1979}
replaced each response $T_{ik}$ with its conditional expectation
$\hat{Y}_{ik}(\beta) = E_{\beta}(T_{ik}| Y_{ik}, \Delta_{ik}, X_{ik})$,
where the expectation is evaluated at regression coefficients $\beta$.
Let $\hat Y_i(\beta) =
\big(\hat Y_{i1}(\beta), \ldots, \hat Y_{iK}(\beta)\big)^{\top}$.
\citet{Jin:Lin:Ying:on:2006} defined
\begin{equation}
  \label{equ:U}
  U_n(\beta, b)= \sum_{i=1}^n\left(X_{i} - \bar{X}\right)^\top\left(\hat{Y}_{i}(b)-X_{i} \beta \right)=0,
\end{equation}
where
$\bar X = \sum_{i=1}^n X_{i} / n$, and $b$ is an initial estimator of $\beta$.
The solution for $U_n(\beta, \beta)$ is the Buckley-James estimator.
The advantage for fixing the initial value $b$ is to avoid solving
for $U_n(\beta, \beta)$ which is neither continuous nor monotone in $\beta$.
Let the $L_n(b)$ be the solution for $U_n(\beta, b)=0$ given $b$.
Then $L_n(b)$ has a closed-form,
\begin{equation}
\label{equ:Lk}
 L_n(b) = \left[ \sum_{i=1}^{n}(X_{i}-\bar{X})^\top (X_{i}-\bar{X}) \right]^{-1}\left[ \sum_{i=1}^{n}(X_{i}-\bar{X})^\top \left(\hat{Y}_{i}(b)-\bar{Y}(b)\right)\right],
\end{equation}
where
$\bar{Y}(b) = \sum_{i=1}^n \hat Y_{i}(b) / n$.
Equation~\eqref{equ:Lk} leads to an iterative algorithm:
$\hat{\beta}^{(m)}_n=L_n(\hat{\beta}^{(m-1)}_n)$, $m \geq 1$.
If the initial estimator $b$ is consistent and asymptotically normal,
$\hat{\beta}^{(m)}_n$ is consistent and asymptotically normal for every $m$.

Although this estimator is consistent, its efficiency might be 
low because it completely ignores the within-cluster dependence.
We next propose to accommodate dependence using the GEE approach,
which covers the estimator of \citet{Jin:Lin:Ying:on:2006} 
as a special case with working independence.

\section{Inference with GEE}
\label{sect:gee}

For a given initial estimator $b$ of $\beta$, we propose an updated
estimator by solving the GEE
\begin{equation}
  \label{equ:gee}
  U_n(\beta, b, \alpha) = \sum_{i = 1}^{n} (X_i - \bar X)^{\top} \Omega_{i}^{-1}\big(\alpha(b)\big) \left(\hat{Y}_{i}(b)-X_{i} \beta \right)=0,
\end{equation}
where $\bar X = \sum_{i=1}^n X_i / n$, and
$\Omega_{i}^{-1}\big(\alpha(b)\big)$ is a $K \times K$ nonsingular working weight matrix
which may involve additional working parameters $\alpha$, which
may depend on $b$.
For given $\alpha$ and $b$, the solution of the GEEs~\eqref{equ:gee}
has a closed-form
\begin{equation}
  \label{equ:Lb}
  L_n(b, \alpha) = \left [ \sum_{i=1}^{n}(X_{i}-\bar{X})^{\top} \Omega_{i}^{-1}\big(\alpha(b)\big) (X_{i}-\bar{X}) \right ]^{-1}\left [ \sum_{i=1}^{n}(X_{i}-\bar{X})^\top \Omega_{i}^{-1}\big(\alpha(b)\big) \left(\hat{Y}_{i}(b)-\bar{Y}(b)\right)\right ].
\end{equation}
This process can be carried out iteratively, summarized as follows.
\begin{enumerate}
\item[1.] 
Obtain an initial estimate $\hat{\beta}^{(0)}_n = b_n$ of $\beta$ and
initialize with $m = 1$.
\item[2.] 
Obtain an estimate $\hat\alpha_n$ of $\alpha$ given $\hat\beta^{(m-1)}_n$,
$\hat\alpha_n(\hat{\beta}_n^{m-1})$.
\item[3.] 
Update with $\hat\beta^{(m)}_n = L_n(\hat\beta^{(m-1)}_n, \hat\alpha_n)$.
\item[4.] 
Increase $m$ by one and repeat 2 and 3 until convergence.
\end{enumerate}

As in \citet{Jin:Lin:Ying:on:2006}, a consistent and asymptotically
normal estimator is important for avoiding convergence problems.
We propose to use the rank-based estimator with Gehan's weight 
from the induced smoothing approach of \citet{John:Stra:indu:2009}.
This estimator has the same asymptotic property as the non-smoothed
version in \citet{Jin:Lin:Wei:Ying:rank:2003}, but can be obtained
with computation ease; its finite sample performance was also reported
to be as well as the best competing methods \citep{John:Stra:indu:2009}.

The GEEs are most efficient when $\Omega_{i}$ is 
chosen to be the covariance matrix of $\hat{Y}_{i}(b)$.
When $\Omega_{i}$'s are the identity matrix (working independence
with all marginal variances the same), our estimator reduces to
the least squares estimator of \citet{Jin:Lin:Ying:on:2006}.
The working covariance matrix $\Omega_{i}$'s are the same when
all clusters have the same size $K$; they only vary with
$i$ when the cluster sizes are not equal.

For convenience, we assume from now on that
$E(\epsilon_{ik}) = 0$, $i = 1, \ldots, n$, $k = 1, \ldots, K$.
This can be achieved by incorporating appropriate columns of ones
in $X_i$, and, hence, adding intercepts in $\beta$.
Our construction of working covariance involves filling element 
$\Omega_{kl}$, for $k, l\in\{1, \cdots, K\}$, 
of the working covariance matrix $\Omega$.
To allow arbitrary number of unique marginal distributions,
let $m_k \in \{1, \ldots, \kappa\}$ be the index of the $k$th
margin among the $\kappa$ unique marginal distributions.
The conditional expectation $\hat{Y}_{ik}(b)$
is computed as
\begin{equation*}
\hat{Y}_{ik}(b) = \Delta_{ik}Y_{ik}+(1-\Delta_{ik})\left [ \frac{\int_{e_{ik}(b)}^{\infty}u \dif \hat{F}_{k, b}(u)}{1 - \hat{F}_{k,b}\left \{ e_{ik}(b) \right \}} + X_{{ik}}^{\top}b \right ],
\end{equation*}
where $e_{ik}(b) = Y_{ik}-X_{{ik}}^{\top}b$ is the right-censored 
error evaluated at $b$, and $\hat{F}_{k, b}$ is the pooled 
Kaplan--Meier estimator of the distribution function $F_{k, b}$ 
from the transformed data $\{ e_{ir}(b), \Delta_{ir}: m_r = m_k \}$,
which share the same margin $m_k$.
Specifically, $\hat{F}_{k, b}$ is 
\begin{equation*}
\hat{F}_{k, b}(t) = 1 - \prod_{1\leq i \leq n, 1 \le r \le K : m_r = m_k, e_{ir} < t}\left( 1-\frac{\Delta_{ir}}{\sum_{j=1}^n\sum_{1 \le l \le K: m_l = m_k} I\left(e_{jl}(b) \geq e_{ir}(b)\right)}\right).
\end{equation*}

To fill the diagonal elements $\Omega_{kk}$, $1 \le k \le K$, evaluate
the conditional second moment of $\epsilon_{ik}(b)$ given the observed data:
\begin{equation}
\label{equ:Vik}
\hat{V}_{ik}(b)= \Delta_{ik} e_{ik}^2(b) + (1 - \Delta_{ik})
\frac{\int_{e_{ik}(b)}^{\infty}u^2 \dif \hat{F}_{k, b}(u)}{1-\hat{F}_{k, b}\left \{ e_{ik}(b) \right \}},
\qquad i = 1, \ldots, n, \quad k = 1, \ldots, K.
\end{equation}
For a given $b$, we fill $\Omega_{kk}$ by an unbiased estimator
of $\Var\big(\epsilon_{ik}(b)\big)$
\begin{equation}
  \label{equ:vhat}
  \hat \Omega_{kk}(b) =  \frac{\sum_{1 \le i \le n, 1 \le r \le K: m_r = m_k} \hat V_{ik}(b)} {n \sum_{1 \le r \le K} I\{m_r = m_k\}}.
\end{equation}

To fill the off-diagonal elements $\Omega_{kl}$, $k \ne l$, define
\begin{equation}
  \label{equ:ehat}
  \hat e_{ik}(b) = \hat Y_{ik}(b) - X_{ik}^{\top} b,
  \qquad i = 1, \ldots, n, \quad k = 1, \ldots, K,
\end{equation}
the conditional expectation of $\epsilon_{ik}(b)$ given the observed data.
Only when $\Delta_{ik} = 1$ is $\hat e_{ik}(b)$ equal to $e_{ik}(b)$.
For a given $b$, we fill $\Omega_{kl}$, $k \ne l$, by
\begin{equation}
  \label{equ:chat}
  \hat{\Omega}_{kl}(b)= \frac{1}{n} \sum_{i=1}^{n} \hat e_{ik}(b) \hat e_{il}(b).
\end{equation}
Because the construction of $\hat e_{ik}(b)$ does not 
involve the dependence between pair $(k,l)$ in cluster $i$, 
$\hat e_{ik}(b) \hat e_{il}(b)$ does not have expectation 
$\Cov\big(\epsilon_{ik}(b), \epsilon_{il}(b)\big)$ unless
$\Delta_{ik} = \Delta_{il} = 1$.
Nevertheless, $\hat\Omega_{kl}(b)$ is still usable for 
its simplicity in constructing working covariance.

Parsimonious working covariance structures such as exchangeable
(EX) or autoregressive with order 1 (AR1) can be imposed.
Parameters $\alpha$ in the working covariance can be estimated with
method of moment estimator $\hat\alpha_n$ based on $\hat\Omega$
as in the non-censored case \citep{Lian:Zege:long:1986}.
When there is no censoring, the working covariance matrix 
$\hat{\Omega}$ converges to the true covariance matrix.
This is no longer true when censoring is present.
Nevertheless, $\hat{\Omega}$, and consequently, $\hat{\alpha}_n$,
still converges to some limit which helps to
improve the efficiency of the GEE estimation.

Extension to unequal cluster sizes as in a 
recurrent event setting is straightforward.
In this case, it is reasonable to assume identical marginal error 
distributions, hence, identical marginal variances.
The working covariance matrix $\Omega_i$ with dimension $K_i\times K_i$
can be constructed with an given estimator $\hat\alpha_n$ for $\alpha$
for a specified working covariance structure.

Under certain regularity conditions, the proposed estimator is consistent
to the true regression coefficients $\beta_0$ and asymptotically normal.
The asymptotic results are summarized in the following theorems,
whose proofs are sketched in the Appendix.
\begin{thm}
\label{thm:cons}
Under conditions A1--A9 in the Appendix, $\hat{\beta}^{(m)}_n$
is a consistent estimator of the true parameter $\beta_0$ for each $m \ge 1$.
\end{thm}
\begin{thm}
\label{thm:norm}
Under conditions A1--A9 in the Appendix, 
$n^{1/2}(\hat{\beta}^{(m)}_n - \beta_0)$
converges in distribution to multivariate normal with mean zero
for each $m \ge 1$.
\end{thm}

The resampling approach developed by \citet{Jin:Lin:Ying:on:2006}
is adapted to estimate the covariance matrix of $\hat{\beta}^{(m)}_n$.
Let $Z_i$, $i = 1, \cdots, n$, be independent and identically
distributed positive random variables, independent of
the observed data, with $E(Z_{i}) = \Var(Z_{i})=1$.
Define
\begin{equation*}
\hat{Y}_{ik}^*(b) = \Delta_{ik} Y_{ik}+(1-\Delta_{ik})\left  [ \frac{\int_{e_{ik}(b)}^{\infty} u \dif \hat{F}_{k, b}^*(u)}{1-\hat{F}_{k, b}^*\left \{ e_{ik}(b) \right \}}+X_{{ik}}^{\top}b \right ],
\end{equation*}
where
\begin{equation*}
\hat{F}^*_{k, b}(t) = 1-\prod_{1 \le i \le n, 1 \le r \le K: m_r=m_k, e_{ir}<t}\left( 1-\frac{Z_i\Delta_{ir}}{\sum_{j=1}^n\sum_{1 \le l \le K : m_l = m_k} Z_iI\left(e_{jl}(b) \geq e_{ir}(b)\right)}\right).
\end{equation*}
Then the multiplier resampling version of equation~\eqref{equ:Lb} has the following form,
\begin{equation*}
L^{*}_n(b, \alpha) = \left [ \sum_{i=1}^{n}Z_{i}(X_{i} - \bar{X}) \Omega^{-1}_i\big(\alpha(b)\big) (X_{i}-\bar{X}) \right]^{-1}\left [  \sum_{i=1}^{n}Z_{i}(X_{i}-\bar{X})\Omega^{-1}_i\big(\alpha(b)\big)\left \{ \hat{Y}_{i}^*(b)- \bar{Y}^*(b)\right \}\right ],
\end{equation*}
where
$\alpha(b)$ is an estimator of working correlation parameter given
regression coefficients evaluated at $b$ and
$\bar{Y}^*(b) = \sum_{i=1}^{n} \hat{Y}^*_{i}(b) / n$.

For a realization of $(Z_1, \ldots, Z_n)$ and an initial estimator
$\hat\beta_n^{(0)}$, a bootstrap estimator of $\beta$ is obtained from 
iteration $\hat{\beta}^{(m)*}_n = L_n^*(\hat{\beta}^{(m-1)*}_n)$.
The covariance matrix of $\hat\beta^{(m)}_n$ can be estimated from the
sample covariance matrix of a bootstrap sample of $\hat\beta^{(m)*}_n$.
The consistency of this variance estimator can be proved following
arguments similar to those in \citet{Jin:Lin:Ying:on:2006}.

\section{Simulation Study}
\label{sect:simu}
We conducted two simulation studies to assess the performance
of proposed estimators and compared its efficiency with 
the initial estimators from \cite{John:Stra:indu:2009}.
The first study had a clustered failure time setting with
identical regression coefficients across margins
and identical marginal error distributions.
The cluster sizes were fixed at three.
For cluster $i$, the multivariate failure time 
$T_i = (T_{i1}, T_{i2}, T_{i3})$ was generated from
\begin{equation*}
\log T_{ik} = 2 + X_{1ik} + X_{2ik} + \epsilon_{ik},
\end{equation*}
where $X_{1ik}$ was Bernoulli with rate 0.5,
$X_{2ik}$ was $N(0, 0.5^2)$, and 
$\epsilon_i = (\epsilon_{i1}, \epsilon_{i2}, \epsilon_{i3})$
was a trivariate random vector specified by identical marginal
error distributions and a copula for the dependence structure.
Three marginal error distributions were considered:
standard normal, standard logistic, and standard Gumbel,
abbreviated by N, L, and G, respectively; the tail of 
the three distributions gets heavier from N to L to G.
The dependence structure was specified by a Clayton copula with three
levels of dependence measured by Kendall's tau: 0, 0.3, and 0.6.
Censoring times were independently generated from uniform
distributions over $(0, c)$, where $c$ was selected for each margin to achieve
three levels of censoring percentage: 0\%, 25\%, and 50\%.
We considered random samples of size $n = 200$ clusters.
Rank-based estimator with Gehan's weight from the induced 
smoothing approach of \citet{John:Stra:indu:2009}, denoted by JS,
was used as the initial estimator for GEE estimators.
Two working covariance structures, EX and AR1, 
were used for the proposed iterative GEE procedure.
The covariance matrix of the estimator was obtained from the
resampling approach with 200 bootstrap size in Section~\ref{sect:gee}.
For each configuration, we did 1000 replicates.

The results are summarized in Table~\ref{tab:seq}.
To save space, only results for nonzero Kendall's tau were reported.
All estimators appear to be virtually unbiased.
The empirical variation of the estimates and the estimated variation
based on the resampling procedure agree closely for all estimators.
For a given censoring percentage, as the dependence level increases,
the variance of the JS estimator changes little, but the variance
of the GEE estimators with both working covariance structures decreases.
Further, the variance from the EX structure is in general smaller
than that from the AR1 structure, which is expected because the true
covariance structure is exchangeable in this simulation setting.
For a fixed dependence level, the effect of censoring percentage on the 
variances of the estimator depends on the marginal error distributions.
The variance increases clearly as the censoring gets heavier when
the errors are normally distributed, but this pattern is not 
observed with Gumbel or logistic marginal error distributions.
The relative efficiency of the proposed GEE estimator in relative
to the rank-based JS estimator is up to 3.5 in the table
(with logistic margin and Kendall's tau 0.6 for $\beta_2$).

\begin{table}[tbp]
\begin{center}
\caption{
Summary of simulation results with identical regression coefficients
and identical marginal error distributions based on 1000 replications.
Empirical SE is the standard deviation of the parameter estimates;
Estimated SE is the mean of the standard error of the estimator;
RE is the empirical relative efficiencies in relative to the JS estimator.
}
\label{tab:seq}
\renewcommand\tabcolsep{3pt}
\begin{tabular}{cccc rrr r rrr r rrr r rr r}
\toprule
Marg & $\tau$ & Cens & $\beta$ & \multicolumn{3}{c}{Bias} &&  \multicolumn{3}{c}{Empirical SE} &&  \multicolumn{3}{c}{Estimated SE} &&  \multicolumn{2}{c}{RE}&\\
\cmidrule(lr){5-7} \cmidrule(lr){9-11} \cmidrule(lr){13-15} \cmidrule(lr){17-19}
&&&& JS & EX & AR1 && JS &EX&AR1&&JS&EX&AR1&&EX&AR1\\
\midrule
N & 0.3 & 0\% & $\beta_1$ & $-$0.002 & $-$0.003 & $-$0.004 &  & 0.087 & 0.072 & 0.075 &  & 0.084 & 0.068 & 0.072 &  & 1.492 & 1.376 \\ 
     &  &  & $\beta_2$ & 0.001 & 0.002 & 0.002 &  & 0.083 & 0.072 & 0.074 &  & 0.084 & 0.068 & 0.071 &  & 1.349 & 1.264 \\ 
     &  & 25\% & $\beta_1$ & $-$0.008 & $-$0.012 & $-$0.013 &  & 0.091 & 0.073 & 0.076 &  & 0.089 & 0.073 & 0.077 &  & 1.543 & 1.415 \\ 
     &  &  & $\beta_2$ & $-$0.003 & $-$0.005 & $-$0.003 &  & 0.093 & 0.075 & 0.079 &  & 0.090 & 0.075 & 0.078 &  & 1.550 & 1.384 \\ 
     &  & 50\% & $\beta_1$ & $-$0.006 & $-$0.011 & $-$0.011 &  & 0.101 & 0.084 & 0.088 &  & 0.099 & 0.086 & 0.090 &  & 1.467 & 1.316 \\ 
     &  &  & $\beta_2$ & $-$0.004 & $-$0.009 & $-$0.010 &  & 0.102 & 0.084 & 0.090 &  & 0.102 & 0.089 & 0.093 &  & 1.484 & 1.281 \\ 
     & 0.6 & 0\% & $\beta_1$ & 0.002 & 0.001 & 0.001 &  & 0.082 & 0.047 & 0.050 &  & 0.083 & 0.046 & 0.050 &  & 3.130 & 2.691 \\ 
     &  &  & $\beta_2$ & 0.005 & 0.001 & 0.001 &  & 0.082 & 0.045 & 0.050 &  & 0.084 & 0.046 & 0.050 &  & 3.316 & 2.697 \\ 
     &  & 25\% & $\beta_1$ & $-$0.007 & $-$0.009 & $-$0.009 &  & 0.092 & 0.050 & 0.055 &  & 0.088 & 0.052 & 0.057 &  & 3.322 & 2.826 \\ 
     &  &  & $\beta_2$ & $-$0.003 & $-$0.008 & $-$0.007 &  & 0.090 & 0.053 & 0.058 &  & 0.090 & 0.054 & 0.058 &  & 2.931 & 2.432 \\ 
     &  & 50\% & $\beta_1$ & $-$0.003 & $-$0.008 & $-$0.008 &  & 0.101 & 0.063 & 0.069 &  & 0.100 & 0.069 & 0.074 &  & 2.567 & 2.144 \\ 
     &  &  & $\beta_2$ & 0.000 & $-$0.005 & $-$0.004 &  & 0.103 & 0.070 & 0.077 &  & 0.102 & 0.071 & 0.077 &  & 2.142 & 1.815 \\ 
L & 0.3 & 0\% & $\beta_1$ & $-$0.001 & 0.002 & 0.004 &  & 0.138 & 0.123 & 0.130 &  & 0.142 & 0.124 & 0.130 &  & 1.258 & 1.128 \\ 
     &  &  & $\beta_2$ & $-$0.006 & $-$0.004 & $-$0.004 &  & 0.145 & 0.125 & 0.130 &  & 0.142 & 0.123 & 0.128 &  & 1.352 & 1.250 \\ 
     &  & 25\% & $\beta_1$ & $-$0.020 & $-$0.022 & $-$0.021 &  & 0.140 & 0.117 & 0.121 &  & 0.145 & 0.121 & 0.128 &  & 1.442 & 1.341 \\ 
     &  &  & $\beta_2$ & $-$0.013 & $-$0.017 & $-$0.018 &  & 0.153 & 0.124 & 0.131 &  & 0.147 & 0.121 & 0.128 &  & 1.512 & 1.369 \\ 
     &  & 50\% & $\beta_1$ & $-$0.011 & $-$0.012 & $-$0.012 &  & 0.164 & 0.133 & 0.140 &  & 0.162 & 0.135 & 0.143 &  & 1.524 & 1.363 \\ 
     &  &  & $\beta_2$ & $-$0.008 & $-$0.013 & $-$0.014 &  & 0.164 & 0.137 & 0.148 &  & 0.166 & 0.137 & 0.145 &  & 1.428 & 1.231 \\ 
     & 0.6 & 0\% & $\beta_1$ & 0.006 & 0.001 & 0.000 &  & 0.145 & 0.084 & 0.093 &  & 0.141 & 0.085 & 0.093 &  & 2.966 & 2.419 \\ 
     &  &  & $\beta_2$ & 0.001 & 0.002 & 0.001 &  & 0.142 & 0.082 & 0.090 &  & 0.142 & 0.085 & 0.092 &  & 3.020 & 2.505 \\ 
     &  & 25\% & $\beta_1$ & $-$0.011 & $-$0.014 & $-$0.015 &  & 0.145 & 0.080 & 0.088 &  & 0.145 & 0.080 & 0.087 &  & 3.245 & 2.679 \\ 
     &  &  & $\beta_2$ & $-$0.014 & $-$0.013 & $-$0.013 &  & 0.149 & 0.080 & 0.088 &  & 0.146 & 0.081 & 0.088 &  & 3.494 & 2.868 \\ 
     &  & 50\% & $\beta_1$ & $-$0.009 & $-$0.011 & $-$0.012 &  & 0.164 & 0.089 & 0.099 &  & 0.162 & 0.094 & 0.102 &  & 3.439 & 2.778 \\ 
     &  &  & $\beta_2$ & $-$0.006 & $-$0.011 & $-$0.012 &  & 0.161 & 0.092 & 0.102 &  & 0.165 & 0.095 & 0.104 &  & 3.036 & 2.479 \\ 
G & 0.3 & 0\% & $\beta_1$ & $-$0.001 & 0.004 & 0.005 &  & 0.092 & 0.092 & 0.096 &  & 0.094 & 0.093 & 0.096 &  & 0.982 & 0.911\\ 
     &  &  & $\beta_2$ & 0.000 & $-$0.004 & $-$0.005 &  & 0.093 & 0.094 & 0.096 &  & 0.094 & 0.093 & 0.096 &  & 0.973 & 0.942\\ 
     &  & 25\% & $\beta_1$ & $-$0.007 & $-$0.015 & $-$0.017 &  & 0.095 & 0.086 & 0.089 &  & 0.093 & 0.085 & 0.088 &  & 1.221 & 1.155 \\ 
     &  &  & $\beta_2$ & $-$0.007 & $-$0.012 & $-$0.014 &  & 0.094 & 0.088 & 0.092 &  & 0.094 & 0.086 & 0.089 &  & 1.140 & 1.048 \\ 
     &  & 50\% & $\beta_1$ & $-$0.008 & $-$0.012 & $-$0.012 &  & 0.099 & 0.089 & 0.091 &  & 0.095 & 0.090 & 0.093 &  & 1.255 & 1.187 \\ 
     &  &  & $\beta_2$ & $-$0.009 & $-$0.013 & $-$0.014 &  & 0.100 & 0.090 & 0.094 &  & 0.097 & 0.092 & 0.095 &  & 1.235 & 1.128 \\ 
     & 0.6 & 0\% & $\beta_1$ & 0.000 & $-$0.004 & $-$0.005 &  & 0.095 & 0.075 & 0.081 &  & 0.094 & 0.072 & 0.077 &  & 1.614 & 1.374 \\ 
     &  &  & $\beta_2$ & 0.001 & $-$0.002 & $-$0.002 &  & 0.094 & 0.074 & 0.079 &  & 0.094 & 0.071 & 0.077 &  & 1.592 & 1.426 \\ 
     &  & 25\% & $\beta_1$ & $-$0.013 & $-$0.015 & $-$0.016 &  & 0.090 & 0.065 & 0.070 &  & 0.093 & 0.065 & 0.070 &  & 1.911 & 1.644 \\ 
     &  &  & $\beta_2$ & $-$0.013 & $-$0.016 & $-$0.015 &  & 0.099 & 0.066 & 0.071 &  & 0.093 & 0.066 & 0.071 &  & 2.231 & 1.918 \\ 
     &  & 50\% & $\beta_1$ & $-$0.012 & $-$0.011 & $-$0.011 &  & 0.093 & 0.069 & 0.074 &  & 0.095 & 0.074 & 0.079 &  & 1.835 & 1.561 \\ 
     &  &  & $\beta_2$ & $-$0.008 & $-$0.013 & $-$0.013 &  & 0.096 & 0.073 & 0.079 &  & 0.097 & 0.077 & 0.083 &  & 1.729 & 1.448 \\ 
  \bottomrule
\end{tabular}
\end{center}
\end{table}

The second simulation setting had multiple event data with different 
regression coefficients and different marginal error distributions.
The cluster sizes were still fixed at three.
For cluster $i$, the multivariate failure times were generated from
\begin{equation*}
\log T_{ik} = \beta_{0k} + \beta_{1k} X_{1ik} + \beta_{2k}X_{2ik} + \epsilon_{ik},
\end{equation*}
where $(\beta_{0k}, \beta_{1k}, \beta_{2k})$, $k = 1, 2, 3$,
was the regression coefficient vector for margin $k$,
and $\epsilon_i = (\epsilon_{i1}, \epsilon_{i2}, \epsilon_{i3})$
was a trivariate random vector specified by three
marginal distributions and a copula for dependence.
The marginal distributions of $\epsilon_i$ were standard
normal, standard logistic, and standard Gumbel, 
respectively, for the first, second and third margin;
their copula was Clayton with three dependence
levels measured by Kendall's tau: 0, 0.3, and 0.6.
The regression coefficients $(\beta_{0k}, \beta_{1k}, \beta_{2k})$
were set to be $(-1, 1, -1)$, $(1, -1, 1)$, and $(1, 1, 1)$,
respectively for $k = 1$, 2, and 3.
Other settings such as the covariates, censoring time, sample size,
initial estimator, bootstrap sample size for variance estimation,
replication size were all the same as in the first simulation setting.
In addition to the JS estimator, GEE estimators with two working 
covariance structures were considered: EX and unstructured (UN).

\begin{table}[tbp]
\begin{center}
\caption{
Summary of simulation results with different regression coefficients
and different marginal error distributions based on 1000 replications.
Empirical SE is the standard deviation of the parameter estimates;
Estimated SE is the mean of the standard error of the estimator;
RE is the empirical relative efficiencies in relative to the JS estimator.}
\label{tab:par}
\renewcommand\tabcolsep{3pt}
\begin{tabular}{ccc rrr r rrr r rrr r rr}
  \hline
&&&\multicolumn{3}{c}{EST}&&\multicolumn{3}{c}{Empirical SE}&&\multicolumn{3}{c}{Estimated SE}&&\multicolumn{2}{c}{RE}\\
\cmidrule(lr){4-6} \cmidrule(lr){8-10} \cmidrule(lr){12-14} \cmidrule(lr){16-17}
  $\tau$ & Cen &$\beta$ & JS & EX & UN && JS & EX & UN && JS & EX & UN && EX&UN\\   
\hline
  0.3 & 0\% & $\beta_{11}$ & 0.008 & 0.003 & 0.003 &  & 0.143 & 0.122 & 0.123 &  & 0.146 & 0.120 & 0.119 &  & 1.370 & 1.351 \\ 
   &  & $\beta_{21}$ & 0.000 & $-$0.003 & $-$0.004 &  & 0.151 & 0.130 & 0.130 &  & 0.146 & 0.120 & 0.119 &  & 1.340 & 1.346 \\ 
   &  & $\beta_{12}$ & $-$0.000 & $-$0.003 & $-$0.002 &  & 0.164 & 0.163 & 0.164 &  & 0.166 & 0.160 & 0.159 &  & 1.014 & 1.006 \\ 
   &  & $\beta_{22}$ & $-$0.001 & $-$0.005 & $-$0.005 &  & 0.162 & 0.160 & 0.161 &  & 0.166 & 0.158 & 0.157 &  & 1.023 & 1.012 \\ 
   &  & $\beta_{13}$ & 0.002 & $-$0.004 & $-$0.003 &  & 0.242 & 0.219 & 0.219 &  & 0.247 & 0.217 & 0.217 &  & 1.221 & 1.220 \\ 
   &  & $\beta_{23}$ & 0.007 & $-$0.001 & $-$0.003 &  & 0.254 & 0.227 & 0.228 &  & 0.249 & 0.217 & 0.217 &  & 1.257 & 1.248 \\ 
   & 25\% & $\beta_{11}$ & 0.008 & 0.004 & 0.003 &  & 0.154 & 0.131 & 0.132 &  & 0.156 & 0.127 & 0.127 &  & 1.374 & 1.368 \\ 
   &  & $\beta_{21}$ & $-$0.005 & $-$0.007 & $-$0.006 &  & 0.160 & 0.132 & 0.132 &  & 0.158 & 0.129 & 0.128 &  & 1.476 & 1.478 \\ 
   &  & $\beta_{12}$ & $-$0.006 & $-$0.001 & $-$0.000 &  & 0.161 & 0.151 & 0.151 &  & 0.165 & 0.148 & 0.147 &  & 1.147 & 1.150 \\ 
   &  & $\beta_{22}$ & $-$0.003 & $-$0.010 & $-$0.010 &  & 0.170 & 0.154 & 0.154 &  & 0.167 & 0.149 & 0.149 &  & 1.217 & 1.209 \\ 
   &  & $\beta_{13}$ & 0.002 & $-$0.006 & $-$0.006 &  & 0.262 & 0.228 & 0.230 &  & 0.260 & 0.220 & 0.219 &  & 1.315 & 1.295 \\ 
   &  & $\beta_{23}$ & $-$0.000 & $-$0.011 & $-$0.012 &  & 0.262 & 0.229 & 0.228 &  & 0.264 & 0.221 & 0.221 &  & 1.310 & 1.321 \\ 
   & 50\% & $\beta_{11}$ & 0.010 & 0.001 & $-$0.000 &  & 0.170 & 0.144 & 0.145 &  & 0.177 & 0.146 & 0.145 &  & 1.381 & 1.376 \\ 
   &  & $\beta_{21}$ & $-$0.018 & $-$0.008 & $-$0.007 &  & 0.180 & 0.150 & 0.150 &  & 0.181 & 0.148 & 0.147 &  & 1.443 & 1.434 \\ 
   &  & $\beta_{12}$ & $-$0.006 & $-$0.005 & $-$0.004 &  & 0.176 & 0.153 & 0.152 &  & 0.169 & 0.149 & 0.148 &  & 1.319 & 1.342 \\ 
   &  & $\beta_{22}$ & 0.014 & 0.004 & 0.002 &  & 0.185 & 0.165 & 0.166 &  & 0.172 & 0.153 & 0.152 &  & 1.261 & 1.241 \\ 
   &  & $\beta_{13}$ & 0.018 & 0.001 & 0.000 &  & 0.315 & 0.270 & 0.271 &  & 0.309 & 0.262 & 0.260 &  & 1.364 & 1.352 \\ 
   &  & $\beta_{23}$ & 0.029 & 0.006 & 0.007 &  & 0.327 & 0.283 & 0.283 &  & 0.314 & 0.264 & 0.262 &  & 1.339 & 1.339 \\ 
  0.6 & 0\% & $\beta_{11}$ & 0.004 & $-$0.000 & $-$0.001 &  & 0.149 & 0.089 & 0.087 &  & 0.146 & 0.084 & 0.092 &  & 2.813 & 2.919 \\ 
   &  & $\beta_{21}$ & $-$0.015 & $-$0.003 & $-$0.002 &  & 0.140 & 0.085 & 0.085 &  & 0.146 & 0.082 & 0.090 &  & 2.700 & 2.722 \\ 
   &  & $\beta_{12}$ & $-$0.010 & 0.000 & $-$0.001 &  & 0.167 & 0.126 & 0.126 &  & 0.165 & 0.120 & 0.142 &  & 1.754 & 1.744 \\ 
   &  & $\beta_{22}$ & $-$0.001 & $-$0.000 & $-$0.000 &  & 0.169 & 0.124 & 0.124 &  & 0.165 & 0.119 & 0.166 &  & 1.873 & 1.853 \\ 
   &  & $\beta_{13}$ & 0.003 & $-$0.004 & $-$0.005 &  & 0.245 & 0.159 & 0.156 &  & 0.248 & 0.156 & 0.192 &  & 2.370 & 2.451 \\ 
   &  & $\beta_{23}$ & $-$0.003 & $-$0.001 & $-$0.000 &  & 0.238 & 0.158 & 0.156 &  & 0.248 & 0.154 & 0.189 &  & 2.279 & 2.326 \\ 
   & 25\% & $\beta_{11}$ & 0.009 & 0.003 & 0.002 &  & 0.155 & 0.093 & 0.092 &  & 0.157 & 0.091 & 0.113 &  & 2.783 & 2.858 \\ 
   &  & $\beta_{21}$ & $-$0.007 & $-$0.004 & $-$0.005 &  & 0.155 & 0.093 & 0.092 &  & 0.159 & 0.093 & 0.112 &  & 2.763 & 2.798 \\ 
   &  & $\beta_{12}$ & 0.000 & $-$0.003 & $-$0.002 &  & 0.166 & 0.113 & 0.113 &  & 0.166 & 0.111 & 0.114 &  & 2.145 & 2.168 \\ 
   &  & $\beta_{22}$ & $-$0.003 & $-$0.006 & $-$0.006 &  & 0.168 & 0.118 & 0.118 &  & 0.167 & 0.112 & 0.114 &  & 2.036 & 2.033 \\ 
   &  & $\beta_{13}$ & 0.001 & 0.000 & 0.000 &  & 0.266 & 0.160 & 0.160 &  & 0.260 & 0.155 & 0.175 &  & 2.769 & 2.771 \\ 
   &  & $\beta_{23}$ & 0.011 & $-$0.000 & 0.000 &  & 0.264 & 0.153 & 0.152 &  & 0.262 & 0.155 & 0.174 &  & 2.991 & 3.028 \\ 
   & 50\% & $\beta_{11}$ & 0.007 & 0.004 & 0.004 &  & 0.174 & 0.112 & 0.111 &  & 0.176 & 0.112 & 0.112 &  & 2.404 & 2.471 \\ 
   &  & $\beta_{21}$ & $-$0.015 & $-$0.005 & $-$0.005 &  & 0.192 & 0.120 & 0.119 &  & 0.179 & 0.118 & 0.117 &  & 2.567 & 2.587 \\ 
   &  & $\beta_{12}$ & $-$0.009 & 0.002 & 0.003 &  & 0.180 & 0.120 & 0.120 &  & 0.169 & 0.119 & 0.120 &  & 2.235 & 2.229 \\ 
   &  & $\beta_{22}$ & 0.017 & 0.005 & 0.003 &  & 0.176 & 0.127 & 0.127 &  & 0.172 & 0.125 & 0.126 &  & 1.911 & 1.923 \\ 
   &  & $\beta_{13}$ & $-$0.000 & $-$0.006 & $-$0.003 &  & 0.307 & 0.199 & 0.196 &  & 0.312 & 0.200 & 0.203 &  & 2.387 & 2.444 \\ 
   &  & $\beta_{23}$ & 0.036 & 0.004 & 0.004 &  & 0.322 & 0.207 & 0.205 &  & 0.315 & 0.204 & 0.203 &  & 2.423 & 2.471 \\ 
         \hline
\end{tabular}
\end{center}
\end{table}

The results are summarized in Tables~\ref{tab:par}.
Similar to the first simulation study, all estimators are
virtually unbiased, and their variance estimators are
generally close to the empirical variances of the replicates.
The variance of the GEE estimators decreases as the 
dependence gets stronger at any level of censoring percentage.
Holding the dependence level, as the censoring percentage
increases, the variance increases at the normal margin, 
but the pattern is different for the other two margins.
The variance has little changes at the logistic margin.
At the Gumbel margin, it remains its level as the censoring 
percentage increases from 0 to 25\%, but increases notably 
as the censoring percentage increases from 25\% to 50\%.
There is almost no difference between the two working 
covariance structures, both leading to about the same relative 
efficiency compared to the rank-based JS estimator.
The relative efficiency of both GEE estimators almost 
double as Kendall's tau is increased from 0.3 to 0.6.

\section{Application}
\label{sect:appl}

The diabetic retinopathy study (DRS) was started in 1971 
\citep{DRS:1976} with the aim to investigate the efficacy of laser 
photocoagulation in delaying the onset of severe vision loss.
Diabetic retinopathy is the most common and serious eye complication 
of diabetes, which may lead to poor vision or even blindness.
A subset of the DRS data for patients with ``high-risk'' diabetic 
retinopathy, categorized by risk group 6 or higher, has been
analyzed by many authors \citep[e.g.,][]{Hust:Broo:Self:mode:1989, 
Lian:Self:Chan:mode:1993, Lee:Wei:Ying:line:1993, Spie:Lin:chec:1996}.
Each of the 197 patients in this subset had one eye randomized 
to laser treatment and the other eye received no treatment.
The outcomes of interest were the actual times from 
initiation of treatment to the time when visual acuity dropped
below 5/200 at two visits in a row (defined as ``blindness'').
The scientific interest was the effectiveness of the 
laser treatment and the influence of other risk factors.
In addition to the treatment indicator, 
three covariates are available:
age at diagnosis of diabetes, 
type of diabetes (1 = adult, 0 = juvenile),
and risk group (6 to 12, rescaled to 0.5 to 1.0).
Since the interaction between treatment and diabetes 
type was found to be significant in \citet{Spie:Lin:chec:1996}, 
we also include this interaction in the model.

\begin{table}[tbp]
\caption{Results of analyzing Diabetic Retinopathy Study.}
\label{tab:dia}
\begin{center}
\begin{tabular}{cc rrr rrr rr}
\toprule
&&\multicolumn{2}{c}{JS}& &\multicolumn{2}{c}{IND}& &\multicolumn{2}{c}{EX}\\
\cmidrule(lr){3-4} \cmidrule(lr){6-7} \cmidrule(lr){9-10}
Margin & Effects & EST & SE && EST & SE && EST &SE\\
\midrule 
\multicolumn{10}{l}{Identical error margins and identical regression coefficients:}\\
pooled& risk group &  $-$2.659 & 0.739 && $-$2.408 & 0.859 && $-$2.306 & 0.775 \\ 
&age&  $-$0.010 & 0.012 && $-$0.010 & 0.013 && $-$0.010 & 0.014 \\ 
&diabetes&  $-$0.140 & 0.349 && $-$0.065 & 0.440 && $-$0.065 & 0.369 \\ 
&treatment&0.520 & 0.197 && 0.545 & 0.330 && 0.542 & 0.263 \\ 
&interaction&  1.116 & 0.301 && 0.961 & 0.466 && 0.964 & 0.410 \\ 
[1ex]
\multicolumn{10}{l}{Different error margins and different regression coefficients:}\\
left  &risk group& $-$2.819 & 1.114 && $-$2.832 & 1.195 && $-$2.654 & 1.242 \\ 
  &age &$-$0.042 & 0.016 && $-$0.037 & 0.019 && $-$0.036 & 0.020 \\ 
  &diabetes & 0.825 & 0.463 && 0.706 & 0.554 && 0.702 & 0.544 \\ 
&treatment& 0.925 & 0.422 && 0.645 & 0.549 && 0.652 & 0.489 \\
 &interaction&1.719 & 0.650 && 1.742 & 0.855 && 1.739 & 0.820 \\
right  &risk group& $-$2.087 & 1.013 && $-$1.944 & 1.316 && $-$1.805 & 1.283 \\
  & age & 0.011 & 0.014 && 0.009 & 0.016 && 0.009 & 0.018 \\ 
  & diabetes & $-$0.770 & 0.432 && $-$0.640 & 0.528 && $-$0.639 & 0.656 \\ 
&  treatment &  0.383 & 0.326 && 0.481 & 0.381 && 0.477 & 0.446 \\  
 &interaction&0.752 & 0.476 && 0.600 & 0.639 && 0.603 & 0.646 \\
[1ex]
\multicolumn{10}{l}{Identical error margins with partial common regression coefficients:}\\
left&age&$-$0.039 & 0.015 && $-$0.036 & 0.021 && $-$0.036 & 0.022 \\ 
 &diabetes &  0.892 & 0.406 && 0.848 & 0.607 && 0.846 & 0.621 \\ 
right &age &  0.011 & 0.015 && 0.009 & 0.019 && 0.009 & 0.017 \\ 
 & diabetes &  $-$0.870 & 0.435 && $-$0.837 & 0.499 && $-$0.835 & 0.574 \\ 
common & treatment &  0.630 & 0.227 && 0.606 & 0.250 && 0.607 & 0.267 \\ 
& risk group &  $-$2.588 & 0.747 && $-$2.409 & 1.034 && $-$2.264 & 0.938\\ 
& interaction &  1.067 & 0.318 && 1.014 & 0.344 && 1.014 & 0.409 \\ 
 \hline
\end{tabular}
\end{center}
\end{table}

\begin{figure}[tbp]
\centering
\includegraphics[width=0.475\textwidth]{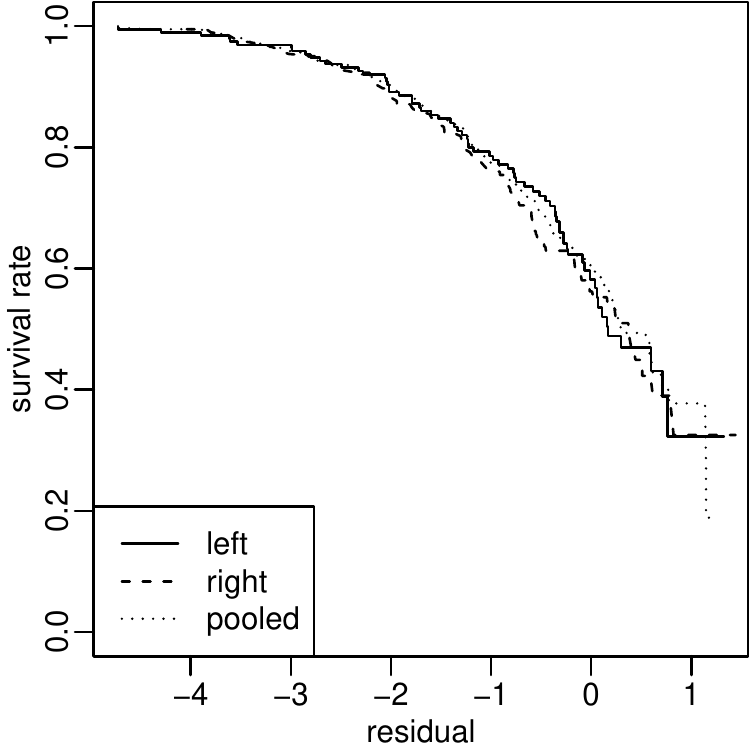}
\includegraphics[width=0.475\textwidth]{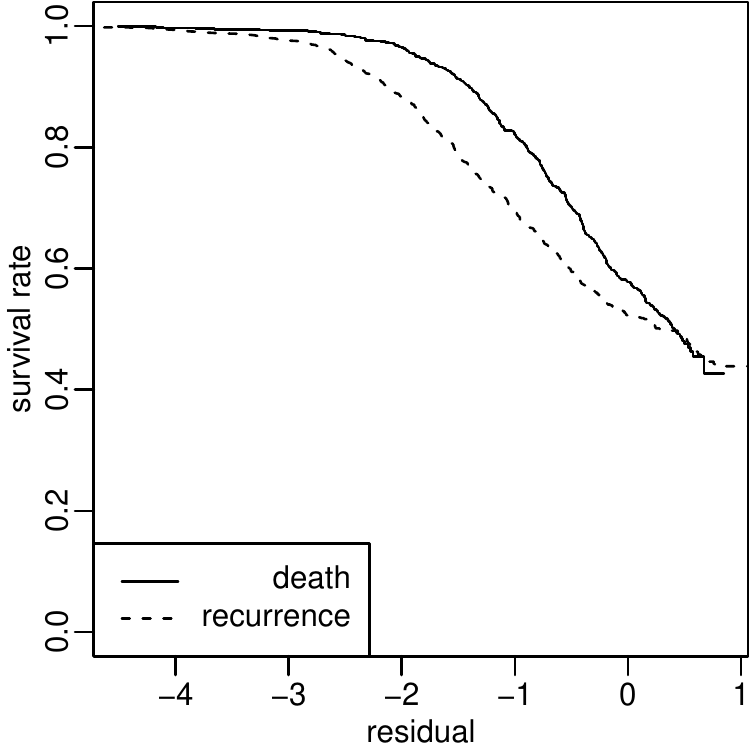}
\caption{Kaplan--Meier survival curves for censored residuals of the
two applications. Left: the DRS Study. Right: the colon cancer study.}
\label{fig:surv}
\end{figure}

We first fit a bivariate AFT model with identical error margins and 
identical regression coefficients for both left and right eyes.
The second AFT model we fit was the opposite, with different error
margins and different regression coefficients for left and right eyes.
For each model, we report GEE estimators with working 
independence and working exchangeable covariance structures, 
in addition to the rank-based JS estimator in Table~\ref{tab:dia}.
The GEE estimator with exchangeable working structure from the
first model suggests that the treatment was significant in 
delaying the onset of vision loss; it had a significant 
higher effect for adult than for juvenile, and patients 
in higher risk groups tended to lose vision sooner.
Note that the treatment effect was not significant if 
working independence were used in the GEE estimator.
The second model offered a possibility to check whether the
marginal error distributions and regression coefficients 
should indeed be identical as assumed in the first model.
Figure~\ref{fig:surv} shows the the Kaplan--Meier survival curves
of the censored residuals for the left margin and right margin
respectively, overlaid with the pooled estimate from the first model.
All three curves appear to be mingled together tightly.
A naive log-rank test to compare the two margins, ignoring that 
the regression coefficients were not known but estimated, yielded
a p-value of 0.907, confirming the visual observation.
Our joint model also allows hypothesis testing of equal coefficients 
for each covariate across the two margins with Wald-type tests.
The coefficients of treatment, risk group, and treatment-diabetes
interaction were found to be not significantly different across
the two margins, with p-values 0.400, 0.278, and 0.147, respectively.
The coefficients of age and diabetes were found to be significantly
different across the two margins, with p-values 0.036 and 0.042, respectively.

We then fit an bivariate AFT model with identical error margins, same 
coefficients for treatment, risk group and treatment-diabetes
interaction, and different coefficients for age and diabetes.
This is one of the many models with intermediate 
complexity between the first model and the second model.
Results are summarized in the last section of Table~\ref{tab:dia}. 
This time, the shared coefficients of treatment, risk group,
and treatment-diabetes interaction remained significant as before.
An interesting finding is that the difference between the coefficient 
of diabetes ($0.846$ versus $-0.835$) is significantly nonzero
with a p-value 0.002, suggesting that the adult diabetes have 
sooner onset of vision loss in right eye than in left eye.
This finding has not been reported in existing analyses.

The second application is a colon cancer study \citep{Lin:cox:1994}.
Through randomization, 315, 310 and 304 patients with stage~C 
colon cancer received observation, levamisole alone (Lev), and 
levamisole combined with fluorouracil (Lev + 5FU), respectively.
\citet{Lin:cox:1994} considered bivariate models for
the time to first recurrence and the time to death.
The research interest was the effectiveness of the treatment
in prolonging the time to recurrence and time to death.
Gender and age are available as covariates besides treatment.

\begin{table}[tbp]
\caption{Result of analyzing Colon Cancer Study}
\label{tab:colo}
\begin{center}
\begin{tabular}{cc rrr rrr rr}
\toprule
&&\multicolumn{2}{c}{JS}& &\multicolumn{2}{c}{EX}\\
\cmidrule(lr){3-4} \cmidrule(lr){6-7}
Margin & Effects & EST & SE & & EST & SE\\
 \midrule
recurrence&  Lev & 0.010 & 0.124 && 0.012 & 0.173 \\ 
 & Lev + 5FU &  0.940 & 0.138 && 0.931 & 0.185 \\ 
&  gender &   0.310 & 0.111 && 0.274 & 0.161 \\ 
&  age &   0.011 & 0.004 & &0.012 & 0.006 \\ 
death &Lev &   $-$0.009 & 0.104 && $-$0.038 & 0.131 \\ 
 & Lev + 5FU &  0.458 & 0.108 && 0.307 & 0.136 \\ 
&  gender &   0.064 & 0.090 && 0.066 & 0.111 \\ 
&  age &  $-$0.003 & 0.004 && $-$0.004 & 0.004 \\ 
   \hline
\end{tabular}
\end{center}
\end{table}

In this application, the error distributions and regression 
coefficients have no reason to be identical across margins.
We report results with different error margin and different
regression coefficients in Table~\ref{tab:colo}.
Since all covariates are at the cluster level, the exchangeable
and independent working covariance structure give the
same results \citep[e.g.,][]{Hin:Care:Wang:crit:2007}.
The Kaplan--Meier survival curves for the two error margins
are shown in Figure~\ref{fig:surv}, which clearly exhibits
no similarity; a naive log-rank test gives p-value $0.0008$.
The treatment of levamisole combined with fluorouracil appears
to have a significant positive effect on both event times.
The gender and age are found not to be significant for either time.
The estimated difference between the combined treatment effect 
on recurrence and on death ($0.931$ versus $0.307$) has a 
standard error $0.103$, suggesting that the combined treatment
has a higher effect on recurrence than on death.

\section{Discussion}
\label{sect:disc}

The working covariance structure of the proposed GEE approach
is different from that in a generalized linear model setting, 
where the variance is assumed to be a function of the mean.
The errors at each margin are assumed to be independent and
identically distributed, and hence have the same variance.
This assumption may be relaxed by imposing a structure
on the variance of the errors.
For instance, in model~\eqref{equ:maft}, we replace 
$\epsilon_{ik}$ with $\sigma_{ik} \nu_{ik}$, where
$\nu_{ik}$'s are independent and identically distributed for 
$i = 1, \ldots, n$ with mean zero and variance one, and the scale 
$\sigma_{ik}$ may be described by a regression model with covariates.
Such specification leads to heteroskedasticity in errors
and merits further investigation.

For applications like the DRS study, where there are reasons
to impose identical distribution across margins, a rigorous test 
to compare the survival curves of the residuals would be desirable.
We used naive tests ignoring the fact that the residuals
were calculated based on estimated regression coefficients.
A rigorous test procedure should take into account of the variation
caused by the estimation procedure.

\par
\appendix
\setcounter{section}{0}%
\setcounter{subsection}{0}%
\renewcommand\thesection       {\Alph{section}}

\section{Sketch of the Proofs}
We impose the following regularity conditions: 
\begin{enumerate}[{A}1:]
\item $\| X_i\| \leq B$ for all $i = 1, \cdots, n$ and some nonrandom constant $B$, where $\|\cdot\|$ is matrix norm. 
\item The density function of $F_{k, \beta}$ exists such that  $\int_{-\infty}^\infty t^2\dif F_{k, \beta}(t) < \infty$, for $k=1, \cdots, K$.
\item The distribution function $F_{k, \beta}$ is twice differentiable with density $f_{k, \beta}$ such that $$\int_{-\infty}^\infty \left( \frac{f_{k, \beta}^\prime(t)}{f_{k, \beta}(t)}\right)^2 \dif F_{k, \beta}(t) < \infty$$ where $1 \leq k \leq K$, and both $f_{k, \beta}(t)$ and $f^\prime_{k, \beta}(t)$ are bounded functions.
\item $E[\exp(\theta\epsilon_{ik}^-)]+ \sup_{k\in\{1, \cdots, K\}} E[\exp(\theta C_{ik}^- )] < \infty$ for some $\theta > 0$, where $a^-=|a|I_{\{a\leq 0\}}$.
\item $\sup_{| b | < \infty; -\infty < t < \infty}\sum_{i=1}^n\sum_{k=1}^K \Pr(t \leq C_{ik} - X_{ik}^\top b \leq t +h) = O(nh)$ as $h \to 0$ and $nh \to \infty$.
\item As $n\to\infty$, $\hat{\alpha}_n$ is bounded and is $n^{1/2}$ consistent to $\alpha_0$ given $\beta$.
\item As $n\to\infty$, initial estimator $b_n$ is $n^{1/2}$ consistent to $\beta_0$ and $\sqrt{n}( b_n - \beta_0)$ is asymptoticly normal with zero mean.
\item The slope matrices $n^{-1} \partial U_n / \partial \beta$ and $n^{-1} \partial U_n / \partial b$ evaluated at $(\beta_0, \beta_0, \alpha_0)$ converge to nondegenerate, finite limit $A$ and $B$, respectively. 
\item The derivative $\partial \Omega_i^{-1}(\alpha) / \partial\alpha$ is finite for all $i = 1, 2, \ldots n$.
\end{enumerate}

Conditions A1--A5 are standard and ensure the existence of 
the solution of equation~\eqref{equ:U} \citep{Lai:Ying:larg:1991}.
It is natural to assume that the working covariance matrix $\Omega$ in 
equation~\eqref{equ:gee} is a symmetric positive definite matrix.
Then there exist a $K\times K$ nonsingular matrix, $\Gamma$, such 
that $\Omega(\alpha_0) = \Gamma^{1/2} \Gamma^{1/2}$.
Let $\mathbb X_i = \Gamma^{-1/2} X_i$, $\mathbb T_i = \Gamma^{-1/2} Y_i$, 
$\mathbb C_i = \Gamma^{-1/2} C_i$, and $\omega_i = \Gamma^{-1/2} \epsilon_i$.
Then equation~\eqref{equ:gee} evaluated at $\alpha = \alpha_0$
can be viewed as equation~\eqref{equ:U} with the transformed 
data $\mathbb X_i$ and $\mathbb Y_i = \min(\mathbb Y_i, \mathbb C_i)$,
with error $\omega_i$, $i = 1, \ldots, n$.
The existence of the solution to equation~\eqref{equ:gee} can be 
verified by the same arguments as in \citet{Lai:Ying:larg:1991},
with assumptions similar to A1 to A5 on the transformed data.
The consistency and asymptotic normality of the estimator 
given $\alpha = \alpha_0$ follow from the same arguments
as in \citet{Jin:Lin:Ying:on:2006}.

The extra complexity here comes from the fact that equation~\eqref{equ:gee}
is solved at $\alpha = \hat\alpha_n$, an estimator of $\alpha_0$.
Under condition A9, the $i$th term in the summation of
$\partial U_n / \partial \alpha$ evaluated at 
$(\beta_0, \beta_0, \alpha_0)$ is a linear function 
of $\hat{Y}_i(\beta_0)-X_i^\top\beta_0$, 
$i = 1, \ldots, n$, with expectation zero.
By the law of large number, $n^{-1}\partial U_n/\partial \alpha$
evaluated at $(\beta_0, \beta_0, \alpha_0)$
converges to zero in probability.

\subsection{Proof of Theorem ~\ref{thm:cons}}
At the solution $\hat\beta_n^{(1)}$ given $b_n$ and $\hat\alpha_n$,
we have $n^{-1} U_n(\hat{\beta}_n^{(1)}, b_n, \hat{\alpha}_n) = 0$.
Taylor expansion at $(\beta_0, \beta_0, \alpha_0)$ gives
\begin{align}
\nonumber
0 =\,& \frac{1}{n}U_n(\beta_0, \beta_0, \alpha_0) + \frac{1}{n} \frac{\partial}{\partial \beta}\left[ U_n(\beta_0, \beta_0, \alpha_0) \right](\hat{\beta}_n^{(1)}-\beta_0) \\
\nonumber
\,& + \frac{1}{n}\frac{\partial}{\partial b}\left[ U_n(\beta_0, \beta_0, \alpha_0) \right](b_n-\beta_0) + \frac{1}{n} \frac{\partial}{\partial \alpha}\left[ U_n(\beta_0, \beta_0, \alpha_0) \right](\hat{\alpha}_n-\alpha_0) + o_p(n^{-1/2}) \\
=\,& \frac{1}{n}U_n(\beta_0, \beta_0, \alpha_0) + A_n (\hat{\beta}_n^{(1)}-\beta_0) +B_n (b_n-\beta_0)+C_n(\hat{\alpha}_n-\alpha_0) + o_p(n^{-1/2}).
\label{equ:taylor}
\end{align}
With regularity conditions A1--A5, the first term converges 
in probability to zero by the law of large number.
The convergence of $b_n$ and $\alpha_n$ in A6 and A7, combined
with the limit condition in A8 and A9, then gives consistency of 
$\hat\beta_n^{(1)}$ to $\beta_0$.
By induction, $\hat{\beta}^{(m)}_n$ is consistent for $\beta_0$ at every $m$.

\subsection{Proof of Theorem ~\ref{thm:norm}}
Under regularity conditions $\sqrt{n}(\hat{\beta}_n^{(1)}-\beta_0)$ 
can be expressed as
\begin{equation}
\sqrt{n}(\hat{\beta}_n^{(1)}-\beta_0) = \left[A_n\right] ^{-1}\left[ \frac{1}{\sqrt{n}} U_n(\beta_0, \beta_0, \alpha_0)+B_n\sqrt{n}(b_n-\beta_0)+C_n\sqrt{n}(\hat{\alpha}_n - \alpha_0)\right] + o_p(1).
\label{equ:app}
\end{equation}
With condition A9, $C_n$ converges to zero in probability, and, hence, with
$\sqrt{n}$ consistency of $\hat\alpha_n$, 
$C_n \sqrt{n} (\hat \alpha_n - \alpha_0) = o_p(1)$.
Equation~\eqref{equ:app} is then asymptotically equivalent to 
\begin{equation*}
\left[A_n\right] ^{-1}\left[ \frac{1}{\sqrt{n}} U_n(\beta_0, \beta_0, \alpha_0)+B_n\sqrt{n}(b_n-\beta_0)\right].
\end{equation*}

With the assumption that $b_n-\beta_0$ is asymptoticly normal, 
there exist some nonrandom functions $\eta_i$ with zero mean such that, 
\begin{equation*}
\sqrt{n}(b_n - \beta_0) = n^{-1/2}\sum_{i=1}^n\eta_i + o_p(\|b_n-\beta_0\|).
\end{equation*}
On the other hand, $U_n(\beta_0, \beta_0, \alpha_0)$ is a 
sum of independent and identically distributed quantities 
with zero mean, denoted by $\phi_i$'s, $i = 1, \ldots, n$.
Equation~\eqref{equ:app} reduces to
\begin{equation*}
\sqrt{n}(\hat{\beta}_n^{(1)}-\beta_0) = \left[A_n\right] ^{-1}\left[ n^{-1/2}\sum_{i=1}^n \left(\phi_i+B_n\eta_i\right)\right] + o_p(\|b_n-\beta_0\|).
\end{equation*}
By multivariate central limit theorem for sums of independent 
random vectors, the asymptotic distribution for $\hat{\beta}_n^{(1)}$ 
is zero mean multivariate normal as $n\to\infty$.
The limit covariance matrix $\Sigma$ have the form 
$A^{-1}\Phi A^{-1}$, where
$\Phi = \lim_{n\to\infty}n^{-1}\sum_{i=1}^n \imath_i \imath_i^{\top}$
with $\imath_i = \phi_i + B\eta_i$.
Induction then implies that $\hat{\beta}_n^{(m)}$ is 
multivariate normal for every $m$.

\bibliographystyle{sinica}
\bibliography{aft}

\begin{thebibliography}{32}
\providecommand{\natexlab}[1]{#1}
\expandafter\ifx\csname urlstyle\endcsname\relax
  \providecommand{\doi}[1]{doi:\discretionary{}{}{}#1}\else
  \providecommand{\doi}{doi:\discretionary{}{}{}\begingroup
  \urlstyle{rm}\Url}\fi

\bibitem[{Brown and Wang(2005)}]{Brow:Wang:stan:2005}
Brown, B.~M. and Wang, Y.-G. (2005).
\newblock Standard errors and covariance matrices for smoothed rank estimators.
\newblock \emph{Biometrika} \textbf{92}, 149--158.

\bibitem[{Brown and Wang(2007)}]{Brow:Wang:indu:2007}
Brown, B.~M. and Wang, Y.-G. (2007).
\newblock Induced smoothing for rank regression with censored survival times.
\newblock \emph{Statistics in Medicine} \textbf{26}, 828--836.

\bibitem[{Buckley and James(1979)}]{Buck:Jame:line:1979}
Buckley, J. and James, I. (1979).
\newblock Linear regression with censored data.
\newblock \emph{Biometrika} \textbf{66}, 429--436.

\bibitem[{Cox(1972)}]{Cox:regr:1972}
Cox, D.~R. (1972).
\newblock Regression models and life-tables (with discussion).
\newblock \emph{Journal of the Royal Statistical Society, Series B,
  Methodological} \textbf{34}, 187--220.

\bibitem[{{Diabetic Retinopathy Study Research Group}(1976)}]{DRS:1976}
{Diabetic Retinopathy Study Research Group} (1976).
\newblock Preliminary report on effects of photocoagulation therapy.
\newblock \emph{American Journal of Ophthalmology} \textbf{81}, 383--396.

\bibitem[{Gehan(1965)}]{Geha:gene:1965}
Gehan, E.~A. (1965).
\newblock A generalized {W}ilcoxon test for comparing arbitrarily
  singly-censored samples.
\newblock \emph{Biometrika} \textbf{52}, 203--223.

\bibitem[{Hin et~al.(2007)Hin, Carey, and Wang}]{Hin:Care:Wang:crit:2007}
Hin, L.-Y., Carey, V.~J., and Wang, Y.-G. (2007).
\newblock Criteria for working correlation structure selection in {GEE}.
\newblock \emph{The American Statistician} \textbf{61}, 360--364.
\newblock \doi{10.1198/000313007X245122}.

\bibitem[{Hornsteiner and Hamerle(1996)}]{Horn:Hame:comb:1996}
Hornsteiner, U. and Hamerle, A. (1996).
\newblock A combined {GEE}/{B}uckley-{J}ames method for estimating an
  accelerated failure time model of multivariate failure times.
\newblock Discussion Paper~47, Ludwig--Maximilians--Universit\"at M\"unchen,
  Collaborative Research Center 386.

\bibitem[{Huang(2002)}]{Huan:cali:2002}
Huang, Y. (2002).
\newblock Calibration regression of censored lifetime medical cost.
\newblock \emph{Journal of the American Statistical Association} \textbf{97},
  318--327.

\bibitem[{Huster et~al.(1989)Huster, Brookmeyer, and
  Self}]{Hust:Broo:Self:mode:1989}
Huster, W.~J., Brookmeyer, R., and Self, S.~G. (1989).
\newblock Modelling paired survival data with covariates.
\newblock \emph{Biometrics} \textbf{45}, 145--156.

\bibitem[{Jin et~al.(2003)Jin, Lin, Wei, and Ying}]{Jin:Lin:Wei:Ying:rank:2003}
Jin, Z., Lin, D.~Y., Wei, L.~J., and Ying, Z. (2003).
\newblock Rank-based inference for the accelerated failure time model.
\newblock \emph{Biometrika} \textbf{90}, 341--353.

\bibitem[{Jin et~al.(2006{\natexlab{a}})Jin, Lin, and
  Ying}]{Jin:Lin:Ying:on:2006}
Jin, Z., Lin, D.~Y., and Ying, Z. (2006{\natexlab{a}}).
\newblock On least-squares regression with censored data.
\newblock \emph{Biometrika} \textbf{93}, 147--161.

\bibitem[{Jin et~al.(2006{\natexlab{b}})Jin, Lin, and
  Ying}]{Jin:Lin:Ying:rank:2006}
Jin, Z., Lin, D.~Y., and Ying, Z. (2006{\natexlab{b}}).
\newblock Rank regression analysis of multivariate failure time data based on
  marginal linear models.
\newblock \emph{Scandinavian Journal of Statistics} \textbf{33}, 1--23.

\bibitem[{Johnson and Strawderman(2009)}]{John:Stra:indu:2009}
Johnson, L.~M. and Strawderman, R.~L. (2009).
\newblock Induced smoothing for the semiparametric accelerated failure time
  model: {A}symptotics and extensions to clustered data.
\newblock \emph{Biometrika} \textbf{96}, 577--590.

\bibitem[{Lai and Ying(1991)}]{Lai:Ying:larg:1991}
Lai, T.~L. and Ying, Z. (1991).
\newblock Large sample theory of a modified {B}uckley-{J}ames estimator for
  regression analysis with censored data.
\newblock \emph{The Annals of Statistics} \textbf{19}, 1370--1402.

\bibitem[{Lee and Wei(1993)}]{Lee:Wei:Ying:line:1993}
Lee, E.~W. and Wei, Z., L. J. aand~Ying (1993).
\newblock Linear regression analysis for highly stratified failure time data.
\newblock \emph{Journal of the American Statistical Association} \textbf{88},
  557--565.

\bibitem[{Li and Yin(2009)}]{Li:Yin:gene:2009}
Li, H. and Yin, G. (2009).
\newblock Generalized method of moments estimation for linear regression with
  clustered failure time data.
\newblock \emph{Biometrika} \textbf{96}, 293--306.

\bibitem[{Liang et~al.(1993)Liang, Self, and Chang}]{Lian:Self:Chan:mode:1993}
Liang, K.-Y., Self, S.~G., and Chang, Y.-C. (1993).
\newblock Modelling marginal hazards in multivariate failure time data.
\newblock \emph{Journal of the Royal Statistical Society, Series B: Statistical
  Methodology} \textbf{55}, 441--453.

\bibitem[{Liang and Zeger(1986)}]{Lian:Zege:long:1986}
Liang, K.-Y. and Zeger, S.~L. (1986).
\newblock Longitudinal data analysis using generalized linear models.
\newblock \emph{Biometrika} \textbf{73}, 13--22.

\bibitem[{Lin(1994)}]{Lin:cox:1994}
Lin, D.~Y. (1994).
\newblock Cox regression analysis of multivariate failure time data: {T}he
  marginal approach.
\newblock \emph{Statistics in Medicine} \textbf{13}, 2233--2247.

\bibitem[{Prentice(1978)}]{Pren:line:1978}
Prentice, R.~L. (1978).
\newblock Linear rank tests with right censored data ({C}orr: {V}70 p304).
\newblock \emph{Biometrika} \textbf{65}, 167--180.

\bibitem[{Qu et~al.(2000)Qu, Lindsay, and Li}]{Qu:Lind:Li:impr:2000}
Qu, A., Lindsay, B.~G., and Li, B. (2000).
\newblock Improving generalised estimating equations using quadratic inference
  functions.
\newblock \emph{Biometrika} \textbf{87}, 823--836.

\bibitem[{Ritov(1990)}]{Rito:esti:1990}
Ritov, Y. (1990).
\newblock Estimation in a linear regression model with censored data.
\newblock \emph{The Annals of Statistics} \textbf{18}, 303--328.

\bibitem[{Robins and Rotnitzky(1992)}]{Robi:Rotn:reco:1992}
Robins, J.~M. and Rotnitzky, A. (1992).
\newblock Recovery of information and adjustment for dependent censoring using
  surrogate markers.
\newblock In Jewell, N., Dietz, K., and Farewell, V. (editors), \emph{AIDS
  Epidemiology --- Methodological Issues}, pages 297--331. Boston, MA:
  Birkh\"auser.

\bibitem[{Spiekerman and Lin(1996)}]{Spie:Lin:chec:1996}
Spiekerman, C.~F. and Lin, D.~Y. (1996).
\newblock Checking the marginal {C}ox model for correlated failure time data.
\newblock \emph{Biometrika} \textbf{83}, 143--156.

\bibitem[{Strawderman(2005)}]{Stra:acce:2005}
Strawderman, R.~L. (2005).
\newblock The accelerated gap times model.
\newblock \emph{Biometrika} \textbf{92}, 647--666.

\bibitem[{Stute(1993)}]{Stut:cons:1993}
Stute, W. (1993).
\newblock Consistent estimation under random censorship when covariables are
  present.
\newblock \emph{Journal of Multivariate Analysis} \textbf{45}, 89--103.

\bibitem[{Stute(1996)}]{Stut:dist:1996}
Stute, W. (1996).
\newblock Distributional convergence under random censorship when covariables
  are present.
\newblock \emph{Scandinavian Journal of Statistics} \textbf{23}, 461--471.

\bibitem[{Tsiatis(1990)}]{Tsia:esti:1990}
Tsiatis, A.~A. (1990).
\newblock Estimating regression parameters using linear rank tests for censored
  data.
\newblock \emph{The Annals of Statistics} \textbf{18}, 354--372.

\bibitem[{Wang and Fu(2011)}]{Wang:Fu:rank:2011}
Wang, Y.-G. and Fu, L. (2011).
\newblock Rank regression for accelerated failure time model with clustered and
  censored data.
\newblock \emph{Computational Statistics and Data Analysis} \textbf{55},
  2334--2343.

\bibitem[{Ying(1993)}]{Ying:larg:1993}
Ying, Z. (1993).
\newblock A large sample study of rank estimation for censored regression data.
\newblock \emph{The Annals of Statistics} \textbf{21}, 76--99.

\bibitem[{Zhou(1992)}]{Zhou:$m$-:1992}
Zhou, M. (1992).
\newblock ${M}$-estimation in censored linear models.
\newblock \emph{Biometrika} \textbf{79}, 837--841.

\end{thebibliography}

\vskip .65cm \noindent Department of Statistics, University of Connecticut, 
215 Glenbrook Rd. U-4120, Storrs, CT 06269, U.S.A.
\vskip 2pt \noindent E-mail: (steven.chiou@uconn.edu and jun.yan@uconn.edu)

\vskip .65cm \noindent Division of Biostatistics, School of Public Health, University of Minnesota, 
A460 Mayo Building, MMC 303, 420 Delaware St., S.E. Minneapolis, MN 55455
\vskip 2pt \noindent E-mail: (junghikim0@gmail.com)

\vskip .65cm \noindent Institute for Public Health Research, University of Connecticut Health Center,
99 Ash Street, 2nd Floor, MC 7160, East Hartford, CT 06108
\vskip 2pt \noindent E-mail: (jun.yan@uconn.edu) \vskip .3cm

\end{document}